\documentstyle[aps,preprint]{revtex}

\begin{document}
\preprint{UCLA/94/TEP/42;
          hep-th/9411110}
\draft
\title{The Life and Times of Emmy Noether\\
{\large Contributions of Emmy Noether to Particle Physics}\footnote{
 Presented at the International Conference on THE HISTORY OF ORIGINAL
IDEAS AND BASIC DISCOVERIES IN PARTICLE PHYSICS, Erice, Italy, 29 July - 4
August 1994. To be published in the Proceedings of the Conference.}}
\author{Nina Byers}
\address{Physics Department, UCLA, Los Angeles, CA 90024}
\date{November 11, 1994}
\maketitle
\begin{abstract}
The contributions of Emmy Noether to particle physics fall into two categories.
One is given under the rubric of Noether's theorem, and the other may be
described as her important contributions to modern
mathematics. In physics literature, the terminology Noether's theorem
is used to refer to one or  another of two theorems, or their converses,
proved by Noether.  These will be discussed along with an historical
account of how they were discovered and what their impact has been.
This paper also gives, for physicists, an overview of the important
role of Emmy Noether's work in the development of modern mathematics.
In addition a brief biography is given.
\end{abstract}

\section*{ DOROTHY HODGKIN 1910-1994}

I would like to dedicate
this lecture to Dorothy Crowfoot Hodgkin, the chemist and x-ray
crystallographer, who died last Friday. She was one of the great scientists of
our century.    Using x-ray  crystallographic methods she got out the chemical
and three-dimensional  structure of complicated molecules, changing organic
chemistry forever and breaking  ground for modern biology.   She was awarded
the Nobel Prize in 1964 for her work, particularly for getting out the
structure of penicillin and  vitamin $B_{12}$ .  But this was only a step along
the way to the pinnacle of her scientific career. In 1969 she got out the
structure of insulin, thirty-five years after she began work on this molecule.
The importance of these discoveries for chemistry, modern biology and  medicine
cannot be overstated.

\section{INTRODUCTION}

Dorothy Hodgkin, Emmy Noether, and an eminent physicist here with us today,
Chien Shiung Wu, are examples of women whose love of science enabled them to
make great contributions in the face of daunting adversity. In  the nineteenth
century women were not admitted into universities and laboratories.  In Germany
and Austria the formal education of women ended at age fourteen. In 1898 the
Academic Senate of the University of Erlangen, where Emmy Noether's father was
professor, declared  the admission of women students would ``overthrow all
academic order.''\cite{kimb}
Women with intellectual interests, born toward the end of
the 19th  century, worked as governesses and language teachers from age
fourteen
until the universities finally admitted women.  Some of these became great
scientists - Marie Curie, Lise Meitner, Emmy Noether.  As the 19th century drew
to a close, the exclusion of women from academic and intellectual life began to
be breached. Then  and in the first  few decades of the 20th century, womens'
colleges  were founded, women were admitted to universities  and  the process
whereby it became possible for women to participate in scientific discovery
began.

In 1900 Emmy Noether was eighteen, and women were finally permitted  to
attend lectures at the University in Erlangen.  They were not allowed to
matriculate, only  to attend lectures - with permission of the instructor. Some
lecturers  refused to lecture if there was a woman in the room.  Emmy Noether
was thus at the forefront of the entry of women into academic life.
One might speculate
whether it is a remarkable accident that a woman of  genius was
among the first, or whether  the social, psychological and emotional
barriers against women doing science  were so formidable that only
women of extremely high ability and determination were able to overcome them.

Emmy Noether was one of the great mathematicians of the 20th century, as all
mathematicians will attest. Not only did she discover the oft quoted theorem
which relates symmetries and  conservation laws, she contributed original
and fundamental ideas to modern mathematics. The importance
of modern mathematical ideas and tools to discoveries in theoretical
elementary particle physics this century is self-evident.

It is fitting therefore that we acknowledge her  contributions at         this
Conference on the  History of Original Ideas and Basic Discoveries in Particle
Physics.   To discuss her contributions to particle physics, it is useful to
separate them into two groups. One is  centered around the theorem we call
Noether's theorem, and the other her seminal contributions to the development
of modern mathematics which has been so influential in theoretical particle
physics. I will discuss the theorem in the next section, its
importance, and the historical context in which it was discovered. The theorem
was published in 1918 and essentially ignored in physics literature for forty
years. There is something of a puzzle as to why it lay fallow for so long since
its relevance to physics, and in particular quantum mechanics, is so clear to
us today. From our modern perspective the theorem reduces the search for
conservation laws and selection rules to the systematic study of the symmetries
of the Lagrangian, and conversely also leads from observed conservation laws to
the discovery of symmetries. In section III is some history which may be
relevant to why  Noether's theorem was so rarely quoted in physics literature
from 1918 to 1958.  Section IV is a   overview for physicists of her original
and highly influential contributions to modern mathematics,  in particular
abstract algebra. Section V is a brief biographical sketch of her life and work
including some details about her father and brother who  were also
mathematicians.  A list of her published papers is in Appendix A; in Appendix B
is her summary of the work published before 1919; in  Appendix C are
72  articles  in
physics and mathematics journals listed in  a recent issue of Current Contents
whose titles
refer   to Noether charges, Noether currents, Noether theorem, etc.
 - of these 26 appear to be in journals of pure mathematics and
the rest in physics and mathematical physics journals!

\section{ THE THEOREM }

 The theorem we so often quote was published in a paper entitled
{\it{Invariante  Variationsprobleme}} in the G\"ottingen Nachrichten  in 1918.
\cite{noet}  It is a very important paper for physics because it proves
very generally
the fundamental relation   of symmetries  and conservation laws.
  The theorem  reduces the search
for conservation laws and selection rules to a  systematic study of the
symmetries of the system, and vice versa, for systems
governed by an action principle whose action  integral is invariant under a
continuous group of symmetry transformations. Noether's paper
combines the theory of Lie groups with the calculus of variations, and
proves two theorems, referred to as I and II, and their
converses. Both theorems and their converses are called Noether's
theorem in the physics literature.   Theorem I pertains to symmetries described
by finite-dimensional Lie groups such as the rotation group, the Lorentz group,
SU(3) or U(1). Theorem II applies for infinite-dimensional Lie groups
such as  gauged U(1) or SU(3) or the group of diffeomorphisms of general
relativity. It is likely that  theorem II was of principal interest  at the
time
the paper was written because, applied to the theory of general relativity,
 from it one obtains energy-momentum conservation as a consequence of the
general coordinate invariance of the theory.  Similarly, and somewhat
more simply, one may obtain
 current conservation as a consequence of gauge invariance in electrodynamics.
Emmy Noether did this work soon after Einstein completed the theory
of relativity  and Hilbert derived the field equations from an action
principle.  Hilbert was concerned by the apparent failure of
`proper' energy conservation laws in the general theory. \cite{klein} It is
characteristic of Emmy Noether   that, having begun this work in response to
Hilbert's questions regarding  energy-momentum conservation in the
general theory, she got results of utmost generality and found theorems that
not only illuminate this question but many other conservation laws as well. I
will describe her two theorems and the historical setting in which they
became known.

Theorem I applies when the symmetry group is a finite-dimensional Lie group;
a Lie group with a finite
number N of infinitesimal generators. Examples are the Lorentz group with
N = 6, and ungauged SU(3) and U(1) with N = 8 and N = 1, respectively.
These generators are the elements
of the Lie algebra. Theorem I states, if the system is invariant with
respect to the Lie group, there is a conserved quantity corresponding to
each element of the Lie algebra.  The result is very general and holds for
 discrete and
continuous, classical and quantum systems. For a field theory, theorem I states
that   there is a locally  conserved current for each element  of the algebra;
i.e., that there are N linearly independent currents $j^{(a)}_{\mu}(x)$
which obey
$\partial^{\mu}j^{(a)}_{\mu} = 0$ where $a$ is a Lie algebra label and $\mu =
0, 1, 2, 3 $ a space-time label. Thus one has
\begin{equation}
$$ \partial^{\mu} j^{(a)}_{\mu} = \partial_0 j^{(a)}_0 +
{\bf{\nabla}}\cdot {\bf{j}}^{(a)} = 0
\end{equation}
for each infinitesimal
generator of the group. From this we obtain  conservation of the
corresponding charge
\begin{equation}
  Q^{(a)} = \int d^3x ~j_0^{(a)}.
\end{equation}
For quantum systems   these charges are operators whose commutation relations
are those of the Lie algebra.

Theorem II applies when the symmetry group is an infinite-dimensional Lie group
(not the limiting case of N $\rightarrow \infty$ for which theorem I continues
to apply). Examples  are the gauged SU(3) and U(1)
groups of QCD and QED,
and the group of general coordinate transformations of general relativity.
Theorem II states that  certain dependencies hold between the
left hand sides of the Euler equations of motion
when the action is invariant with respect
to an infinite-dimensional Lie group.
In the case of general relativity, using Hilbert's Lagrangian
and the invariance of the action under general coordinate transformations,
the dependencies of theorem II are Bianchi identities. The four
Bianchi identities $G_{\mu\nu;\mu}$ = 0 give the energy-momentum conservation
law, as can be seen from the following. For Einstein gravity coupled to
electromagnetism and/or matter, the
field equations are $G_{\mu\nu} = -8 \pi \kappa T_{\mu\nu}$ where $T_{\mu\nu}$
is the energy-momentum tensor of the electromagnetic and/or matter fields.
Theorem II thus gives $T_{\mu\nu;\mu}$ = 0,  which is the
law of energy-momentum conservation in the general theory. Similarly, for QED
theorem II gives current conservation as a consequence of  gauge
invariance.

For the mixed case where the symmetry group is
the union of a finite-dimensional and an infinite-dimensional Lie group,
Noether found both types of results; i.e., conservation laws and dependencies.

 The paper was submitted to the University of G\"ottingen in 1919  as her
Habilitation thesis. Actually Hilbert had tried to  obtain a university
Habilitation for Noether in 1915 when she came to G\"ottingen.
Consideration was refused by the academic senate on grounds she was a woman,
and Hilbert uttered his famous quote `` I don't see why the sex of
the candidate is relevant - this is afterall an academic institution not a
bath house.'' The Habilitation was granted in 1919.
 It is interesting to read how she describes her
results in her submission. She  says  the paper  ``  deals with arbitrary
finite- or infinite-dimensional continuous groups, in the sense of Lie, and
discloses the consequences when a variational problem is invariant under
such a group.  The general results contain, as  special cases, the
theorems on first integrals as they are known in mechanics and,  furthermore,
the conservation theorems and the dependencies among the field equations
in the theory of relativity -- while on the other hand, the converse of these
theorems is also given ...''\cite{Dick}  In the Abstract to the paper she
wrote
``The variational problems  here considered are such as to admit a  continuous
group (in Lie's sense); the conclusions that emerge for the differential
equations find their most general expression in the theorems formulated   in
section I and proved in the following sections. Concerning differential
equations that arise from problems of variation, far more precise statements
can be made than about arbitrary differential equations admitting a group,
which are the subject of Lie's researches.  For special groups and
variational problems,  this combination of methods is not new; I may
cite Hamel and
Herglotz for special finite-dimensional groups, Lorentz and his pupils ( e.g.,
Fokker,  Weyl and Klein ) for special infinite-dimensional groups. In
particular Klein's second Note and the present developments have been mutually
influenced by each other. In this regard I refer to the concluding remarks of
Klein's Note.'' \cite{tavel}

The Note of Klein she refers to is entitled {\it{\"Uber die Differentialgesetze
f\"ur die Erhaltung von Impuls and Energie in der Einsteinschen
Gravitationstheorie.}}  It ends with an acknowledgement to Noether saying
``I must not fail  to thank Frl. Noether again for her valuable participation
in my new work... Her general treatment is given in these  Nachrichten in a
following Note.'' His work was presented to the Gesellschaft der
Wissenschaften at a 19 July 1918 meeting, and he
says he presented her more general results the following week.\cite{felix}

Noether's  interest in the general theory was somewhat aside from the main
path of her mathematical research as reflected in her publication list
(Appendix A) but very understandable since she came to work in G\"ottingen in
1915 at Hilbert's invitation, and Hilbert says he asked her to look into the
question of energy conservation in Einstein's theory. \cite{klein} G\"ottingen,
 at
that time, was  the world center of mathematics; Hilbert had assembled there a
stellar array of mathematicians.  Felix Klein, Hermann Minkowski, and Karl
Schwarzschild were among them. There was intense interest in the general theory
of relativity. Hermann Weyl  said ``Hilbert was over head and ears  in the
general theory, and for Klein the theory and its connection with his old ideas
of the Erlangen Program  brought the last flareup of his mathematical interests
and production.'' \cite{weyl}  Noether published two papers directly relating
to
 the
general theory - No. 12 and 13 in the her publication list. Weyl characterized
these papers as follows.  ``For two of the most significant aspects of general
relativity theory she gave the correct and universal mathematical formulation:
first, the reduction of the problem of differential invariants to a purely
algebraic one by the use of `normal coordinates'; and second, identities
between the left sides of Euler's equations of a  variational problem which
occur when the  [ action ] integral is invariant   with respect  to a group of
transformations involving arbitrary functions.'' The paper we are considering
here is the second of these two. It is interesting to examine further the
historical situation. In the summer of 1915 Einstein gave six lectures in
G\"ottingen on generalizing the  special theory of relativity to include
gravity.  At this time, according to Pais \cite{pais}, he did not yet have
the theory
completed but he  felt he had \lq... succeeded to convince Hilbert and Klein
.....\rq. In the fall, Einstein  found, at last, found the correct field
equations.   At the same time Hilbert also got the same equations by writing a
Lagrangian  for  the theory and deriving the field  equations from an action
principle. Weyl was very impressed, as was everyone, and he quickly wrote his
book  {\it{Raum - Zeit - Materie}}. The first edition
was published in 1918.  It begins `` Einstein's theory of relativity has
advanced our ideas of the structure of the cosmos a step further. It is as if a
wall which separated us from truth has collapsed.''  \cite{stm}

Hilbert wrote an article entitled {\it{Grundlagen der Physik}} and remarked
there on the failure in the general theory of ordinary laws of energy-
momentum conservation; Klein published a
correspondence with Hilbert on this. \cite{klein}
Proof of local energy conservation is
not clear as it is in
Newtonian theories. The conservation laws in those theories were called
``proper'' by Hilbert and he found that they failed in the general theory.
After proving that the by now accepted form of the energy-momentum conservation
law follows from the invariance of the theory under general coordinate
transformations, Noether concludes her paper with a section entitled A
HILBERTIAN ASSERTION that begins `` From the foregoing, finally, we also obtain
the proof of a Hilbertian assertion about the connection of the failure of
proper laws of conservation of energy  with  general relativity, and prove this
assertion in a generalized group theory version.''  She proved that generally
one has  what they then called improper  energy relationships when the symmetry
group is an infinite-dimensional  Lie group, and in addition to the general
theory of relativity she gave another example of this. It was her style,
starting from a specific case,  to get the most general results.

The generality of her results is characteristic of  the whole body of her work.
The overall distinguishing characteristic of her major contributions to modern
mathematics stems from her  ability to abstract matters of general importance
from details.   According to her student van der Waerden, her work was guided
by a maxim he described as follows.\cite{van} ``Any  relationships between
 numbers,
functions and operations only become transparent, generally applicable, and
fully productive after they have been isolated from their particular objects
and been formulated as universally valid concepts. Her originality lay in the
fundamental structure of her creative mind, in the mode of her thinking and in
the aim of her endeavors.  Her aim was directed specifically towards scientific
insight.''

In the 1916 - 1918 period her work was widely recognized. Einstein wrote to
Hilbert in the spring of 1918 `` Yesterday I  received from Miss Noether a
very interesting paper on invariant forms. I am impressed that one can
comprehend these matters from so general a viewpoint.  It would not have done
the Old Guard at G\"ottingen any harm had they picked up  a thing or two from
her. She certainly knows what she is doing.'' \cite{kimb} He is
probably refering to No. 12 in Appendix A.  This is the only one of
Noether's papers  cited in in Pauli's 1921 {\it{Encyklopaedie der
mathematischen Wissenschaften}} article on relativity.  It seems odd that
her didn't also reference No. 13. Perhaps this was a harbinger of
things to come.  In the twenties and thirties, and indeed for
about forty years, Noether was rarely cited in the literature
though her results were given often.  It is not clear why this is so. Perhaps
it is because of an ambiguity  having to do with Klein's Note.
In the acknowledgement he makes to her contributions to his work,
there is perhaps some  insinuation that
he was somehow responsible for her results.\cite{felix}

There is no paucity of references to Noether's theorems in contemporary
literature.   As regards theorem II, Peter G. Bergmann wrote in 1968:
``Noether's theorem forms one of the cornerstones of work in general
relativity. General relativity is characterized by the principle of general
covariance according to which the laws of nature are invariant with respect to
arbitrary curvilinear coordinate transformations that satisfy minimal
conditions of continuity and differentiability. A discussion of the
consequences in terms of Noether's theorem would have to include all of the
work on ponderomotive laws, ... ''\cite{kimb}  Feza  Gursey wrote
in 1983:  ``The key to the relation of symmetry laws to conservation laws is
Emmy Noether's celebrated theorem. ...  Before Noether's theorem the principle
of  conservation of energy was shrouded in mystery, leading to the obscure
physical systems of Mach and Ostwald.  Noether's simple and profound
mathematical formulation did much to demystify physics. ... Since all the laws
of fundamental physics can be expressed in terms  of quantum fields which are
associated with symmetry groups at each point and satisfy differential
equations derived from an action principle, the conservation laws of physics
and the algebra of time-dependent charges can all be constructed using
Noether's methods.  The only additional conserved quantities not connected with
the Lie algebra are topological invariants that are related to the global
properties of the fields.  These have also become important in the last few
years.  With this exception, Noether's work is of paramount importance to
physics and the interpretation of fundamental laws in terms of group theory.''

Now Noether's theorem  is  a basic tool in the arsenal of the theorist,  and
is taught in every class on quantum field theory and particle physics. It is
curious that  it  seems to have lain fallow in the physics literature  for
nearly forty years being mentioned very rarely from 1920 to 1960.   In the next
section are some further comments and conjectures regarding this.

\section{ A PUZZLE}

The puzzle is why were there so few references to E. Noether in the physics
literature for nearly forty years?\cite{kast}
Now her name appears very frequently,
and  most textbooks
on classical and quantum mechanics and
classical and quantum field theory
have sections entitled Noether's theorem.
Actually her results did not fall into obscurity but they were often given
 without a reference to her.  This may have begun with Hermann Weyl's
 important book {\it{ Raum - Zeit - Materie}} in which
 he derives the energy-momentum conservation law for relativity from
general coordinate invariance. He does not refer to Klein or Noether in
the text. In  a footnote  he
references  the Klein  paper, and adds ``Cf., in the same periodical,
the general formulations given by E. Noether.'' The English version in
which one finds this is a translation of a 1919 edition. \cite{stm}
 In the first edition, dated
Easter 1918 in Mecklenburg, he gets an energy-momentum conservation law
from the field equations. Obviously he was not aware then of  Noether's
theorem and Klein's Note. In the preface to the 1919 edition he says
  ``Chapter IV, which is in the main devoted
to Einstein's theory of gravitation, has been subjected to a very considerable
revision in consideration of the various important works that have appeared,
in particular those that refer to the Principle of Energy-momentum.''
Perhaps because Weyl's book was very important,
and he did not mention Noether's theorem, others followed suit.

There is one important book written in the twenties that mentions
Noether's theorem. A short subsection devoted to Noether's
theorem is in Courant and Hilbert's
{\it{Methods of Mathematical Physics}}; the German edition
 was first published in 1924. \cite{hilb}

Perhaps a more substantial
reason for the paucity of references to Noether's theorem
 in the twenties and thirties, than that Weyl didn't mention it,
is that her theorems  were not felt to be of fundamental importance.
In that period,
energy conservation and general relativity  were not as firmly established
as they are now. \cite{bram}  Of course no one
doubted macroscopic energy conservation; the first law of
thermodynamics had been firmly established by 1850.  But the discovery
of radioactivity, particularly the continuous $\beta$ spectrum, raised serious
doubts regarding energy conservation as a fundamental principle. Though
Chadwick had presented evidence of a continuous $\beta$ spectrum in 1914,
his results were not definitive and some thought that the
electrons were monoenergetic and the observed continuous $\beta$ spectrum an
experimental artifact.
Lise Meitner was among those who
believed that energy conservation was a fundamental principle and that
there must be narrow lines underlying the $\beta$ spectrum.  It was only in
1927
that Chadwick and Ellis gave convincing evidence in the form of calorimetric
measurements that the $\beta$ spectrum was continuous. Meitner
then  confirmed those results in her own laboratory, and this
 provoked Pauli's proposal  of the neutrino in December 1929. \cite{meit}
  Though energy-momentum conservation  had been clearly demonstrated
experimentally in Compton scattering in 1925, Pauli's neutrino hypothesis did
not immediately reinstate energy conservation as a fundamental principle.
For example,  Bohr proposed energy nonconservation in
nuclear processes in his Faraday lecture at Caltech in 1930. He wrote to
Mott in October 1929 ``I am preparing an account on statistics and
conservation in quantum mechanics in which I also hope to give convincing
arguments for the view that the problem of $\beta$-ray expulsion lies
outside the reach of the classical conservation principles of energy
and momentum.'' \cite{bram}   Pais writes that Bohr continued to consider the
possibility that energy is not conserved in $\beta$-decay until 1936.
You might think that   Fermi's incorporation of Pauli's neutrino
hypothesis in his theory of beta decay (published December 1933)
would have reestablished  the credibility of energy conservation as a
fundamental principle. However,  in
1936 there were experimental indications (later proved false) of failure of
the conservation laws in the Compton effect and, for example, Dirac wrote a
paper entitled \lq Does conservation of energy hold in atomic processes?'
\cite{dirac} It was not until 1939 that measurements of $\beta$-spectra in
allowed transitions confirmed Fermi's theory. \cite{laws}
Energy conservation in atomic processes was not in doubt, at least not for
long;
but it does seem that energy conservation as a fundamental principle
was in doubt. Perhaps at some level it remained so all the way
until 1956, when the definitive experimental verification of
energy conservation in $\beta$ decay was achieved with the
direct detection of $\nu_e$ by Reines and Cowan. \cite{cowa}

With the advent of quantum mechanics, one might have thought that Noether's
theorem would have been invoked.   The connection between
symmetries and conservation laws was of fundamental interest. Nevertheless
it is remarkable that the only reference to Emmy Noether in  Weyl's
{\it{Theory of Groups and Quantum Mechanics}} is to her paper
generalizing the Jordan-H\"older theorem.  \cite{tgqm}  He uses her
theorem II in his treatment of  the Dirac electron
in interaction with the  electromagnetic field, but without reference
to her paper. From gauge invariance of
the action, he obtains conservation of current and then
 shortly thereafter says
``Just as the theorem of conservation of electricity follows from the
gauge invariance, the theorems for conservation of energy
and momentum follow from the circumstance that the action integral, formulated
as in the general theory of relativity, is invariant under arbitrary
(infinitesimal) transformations of  coordinates.''   Perhaps
he omitted referencing her 1918 paper because by the time his book
on group theory and quantum mechanics was
written (1928), her more recent work overshadowed, for mathematicians,
her theorems on symmetries and conservation laws.
Nevertheless, she might have benefited from multiple citations of her
work.    Her status in the University was far below what she
merited on the basis of her accomplishments and ability. Weyl had been a
visitor to G\"ottingen in 1926-27.  In the
address he gave at her memorial service he said ``When I was called
permanently to G\"ottingen in 1930, I earnestly tried to obtain a better
position for her, because I was ashamed to occupy such a preferred position
beside her whom I knew to be my superior as a mathematician in many respects.
I did not succeed... Tradition, prejudice, external considerations weighted
the balance against her scientific merits and scientific greatness, by that
time denied by no one.  In my G\"ottingen years, 1930-1933, she was without
doubt the strongest center of mathematical activity there, considering both
the fertility of her scientific research program and her influence upon a
large circle of pupils.''\cite{weyl}

Perhaps another reason Noether's theorem was not given much publicity was
because it may have felt awkward for pre-WWII authors to have credited a
woman for an important contribution to their work.

{}From a contemporary perspective it seems surprising that Weyl did not
use Noether's theorem I to obtain, for example, conservation of
angular momentum from rotational invariance. This, however, doesn't
fit into the approach of his book because he
uses a Hamiltonian rather than a Lagrangian formulation of quantum
mechanics.

In the 1950's when Lagrangian formulations became more prevalent,
references to Noether's theorem began appear in the literature.
Kastrup describes the major papers that seem to bring it forward.\cite{kast}
The first quantum field theory text I have found that mentions Noether's
theorem is Bogoliubov and Shirkov's {\it{Introduction to
the Theory of Quantized Fields}}. \cite{bogo} This book
presents classical and quantum field theories from a
Lagrangian point of view, and devotes a subsection to Noether's
theorem (theorem I) in what is essentially the first chapter.
Gregor Wentzel's
book {\it{Quantum Theory of Fields}}, widely used in the forties and fifties,
does not use it or refer to it, though  in a footnote to the
section entitled Conservation Laws
he remarks that ``the validity of the conservation laws is known to
be  connected with certain invariance properties of the
Hamiltonian.''\cite{went}
In the text he derives energy-momentum conservation when the
Hamiltonian does not depend upon
space-time  coordinates by  construction of a divergence-free symmetric
energy-momentum tensor using the field equations.  His book generally
gives a Hamiltonian rather than a Lagrangian formulation of quantum
field theory.  In the footnote mentioned above, Wentzel references
Pauli and Heisenberg \cite{paul}. In their famous papers on quantum field
theory,
there is no reference to Noether.  Theirs is also
a Hamiltonian approach to the subject.

 The frequency with which Noether's theorem is referred to
in physics literature, particularly particle physics literature,
 increased substantially after 1958.
 This was the year that the
Feynman and Gell-Mann
paper on the V-A theory of weak interactions was published.
\cite{feyn} Though no reference is made to Noether's theorem,
 Feynman and Gell-Mann
clearly point to the  connection of conserved currents and
symmetries.  They propose in that paper the  conserved vector
 current (CVC) hypothesis,
observing that  the  decay rates of the muon
and $O^{14}$ give nearly equal values for the
Fermi coupling  constant. From this observation they suggest that
the Fermi coupling constant may be a weak charge related to the conserved
weak vector current as in (2). Like electric charge,
it appears that it is not renormalized by the strong interactions and
is the same for leptons and hadrons.
Probably with reference to conservation of the electromagnetic current
as a consequence of gauge invariance, they presciently seem to be suggesting
that another gauge principle may be involved; the  final sentence of their
paper reads in part:
``it may be fruitful to analyze further the idea that the
vector part of the weak coupling is not renormalized; ... and to study the
meaning of the transformation groups which are involved.''
Another paper that was influential at around the same time was Schwinger's 1957
Annals of Physics paper {\it{A Theory of the
Fundamental Interactions}}.\cite{schw}  In his theory,
internal symmetries are described by finite-dimensional Lie groups
and  he uses, without
reference to Noether, her theorem I.  Indeed it plays an important role
in his theory.

It seems to me that
these papers along with  the coming back into vogue of Lagrangian
field theory, led people to feel that Noether's theorem was important or,
anyway, useful.
Previously, and to some extent still at that time,
people used a Hamiltonian approach for theoretical elementary particle
physics even
though Schwinger's formulation of quantum field theory
in terms of an action principle had been enormously
influential. Up until this time and even a bit
beyond, most theorists were not thinking of theories of strong
or weak interactions as Lagrangian field theories governed by
an action principle. \cite{gross}
Schwinger's 1957 paper is somewhat exceptional in
this respect and perhaps, for some at least, led the way. As long as
 Lagrangian field theory was not seen as the
starting point for a theory of elementary particles, Noether's theorem was
 not as consequential
as it later became.  Later when theorists began to use path integrals,
Lie groups, and gauge symmetries, Noether's
theorem  became a basic tool in their arsenal.

It may be an amusing coincidence that two of the possible roadblocks to
frequent mention of Noether's theorem in the older literature
disappeared at about the same time.
Final confirmation of the principle of
energy conservation  by Reines and Cowan's direct detection
of $\nu_e$  occured at roughly the same
time as widespread recognition of the
importance of Lie groups in  Lagrangian formulations  of quantum field
theories began.

Perhaps we will learn that energy-momentum conservation is not a  fundamental
principle after all; i.e., that the diffeomorphism symmetry of space-time is
violated at small distances. Nevertheless Noether's theorem will remain an
important  contribution to physics because it gives, in general, the relation
between conservation laws and symmetries. Furthermore the theorem formulated by
Noether with such depth and generality  has contributed very  importantly  to
modern physics both in the discoveries of symmetries  of fundamental
interactions and in finding the dynamical consequences of symmetries. I believe
I would not be alone in asserting that her theorems  have played a key role in
the development of theoretical physics.

\section{ CONTRIBUTIONS BEYOND THE THEOREM}

     As important as the theorem is, it
by no means sums up her contributions to modern physics. From her
point of view, and that of her mathematical colleagues, the two 1918
papers constituted a tangent to a main road of accomplishment. This
road was to establish modern abstract algebra.
It is self-evident that modern mathematics is, and has been, a very
important contributor to discovery in particle physics. Modern abstract
algebra  profoundly affected
modern mathematics in general; to quote  Michael Atiyah
`` Modern mathematics, in all its branches, has been
influenced by a more liberal and ambitious use of algebra.  In recent
years this is also increasingly true of theoretical physics.  Lie groups,
commutation relations, supersymmetry, cohomology, and representation theory
are widely used in theoretical models for particle physics. Emmy Noether's
belief in the power of abstract algebra has been amply justified.'' \cite{atiy}
Nathan Jacobson wrote in the introduction to her collected
works  that
``Emmy Noether was one of the most influential mathematicians of this century.
The development of abstract algebra, which is one of the most distinctive
innovations of  twentieth century mathematics, is largely due to her - in
published papers, in lectures, and in personal influence on her
contemporaries.'' \cite{jacob}
Concepts, methods and results in
group theory, algebraic topology, cohomology theory,
homotopy theory, etc.  are
valuable tools for understanding physics. To give some recent examples,
 methods and concepts from algebraic topology are very usefully employed in
analytic studies of gauge field theories on the lattice \cite{tomb}; and
higher
homotopy groups are found useful in analyzing
possible forms of spontaneous symmetry breaking. \cite{dhok}

 In this section I will  give a brief overall summary of Emmy Noether's
contributions drawn principally from writings of Weyl \cite{weyl},
Jacobson \cite{jacob} and van
der Waerden  \cite{van}. Since we are not mathematicians, it is difficult to
 give
here a complete and accurate account of her major contributions.
A list of her published papers is given in Apppendix A.
Hermann Weyl said, however, that  ``one
cannot read the scope of her accomplishments from individual results of her
papers alone; she originated above all a new and epoch-making style of thinking
in algebra.''   He writes about
her work as follows. ``Emmy Noether's scientific production fell into three
clearly distinct epochs; (1) the period of relative dependence, 1907-1919; (2)
the investigations grouped around the general theory of ideals 1920-1926; (3)
the study of the non-commutative algebras, their representations by linear
transformations, and their application to the study of commutative number
fields and their arithmetics.''  As regards the first epoch, I have
already written
about the 1918 papers; the other dozen or so papers show
her thinking developing from the old ( 19th century ) ways of doing algebra and
invariant theory to the new ideas of what Weyl calls the second epoch. She
summarized her work in that first epoch in her Habilitation submission.  I
have included that here as Appendix B.

To summarize her work after 1919,
let me begin by quoting the Russian
topologist P. S. Alexandrov. ``When we speak of Emmy Noether as a
mathematician we mean not so much the early works but instead the period
beginning about 1920 when she struck the way into a new kind of algebra.
.....[She] herself is partly responsible for the fact that her work of the
early
period is rarely given the attention  [among mathematicians]  that it would
naturally deserve: with the singlemindedness that was part of her nature, she
herself was ready to forget what she had done in the early years of her
scientific life, since she considered those results to have been a diversion
from the main path of her research, which was the creation of a general,
abstract algebra. It was she who taught us to think in terms of simple and
general algebraic concepts - homomorphic mappings, groups and rings with
operators, ideals ... theorems such as the `homomorphism and isomorphism
theorems', concepts such as the ascending and descending chain conditions for
subgroups and ideals, or the notion of groups with operators were first
introduced by Emmy Noether and have entered into the daily practice of a wide
range of mathematical disciplines. ...  We need only glance at Pontryagin's
work on the theory of continuous groups, the recent work of Kolmogorov on the
combinatorial topology of locally compact spaces, the work of Hopf on the
theory of continuous mappings, to say nothing of van der Waerden's work on
algebraic geometry, in order to sense the influence of Emmy Noether's ideas.
This influence is also keenly felt in H. Weyl's book  {\it{Gruppentheorie und
Quantenmechanik.}}'' \cite{alex}  All who have written about her recall that
she
always worked with a lively group of mathematicians around her.
She gave lecture courses  in G\"ottingen and elsewhere and loved to
talk mathematics with groups of like-minded mathematicians. She had many
very good students \cite{Dick} and her influence
extended well beyond her published papers. A notable example is given by
Jacobson. `` As is quite well known, it was Noether who persuaded P.S.
Alexandrov and Heinz Hopf to introduce group theory into combinatorial topology
and formulate the then existing simplical homology theory in
group-theoretic terms in place of the more concrete setting of incidence
matrices.''  Alexandrov and Hopf say in the preface to their book Topologie
(Berlin 1935) ``Emmy Noether's general mathematical insights were not confined
to her specialty - algebra - but affected anyone who came in touch with her
work.''

It was in the second epoch  according to Weyl, 1920-26, that she founded
the approach of modern abstract algebra.  Jacobson describes how this
came about; numbers refer to the list in Appendix A.
``Abstract algebra can be dated from the publication of two papers by
Noether, the first, a joint paper with Schmeidler, {\it{Moduln in
nichtkommutativen Bereichen ...}} (no.17) and
{\it{Idealtheorie in Ringbereichen}} (no.19). Of these papers, ..., the first
is of somewhat specialized interest and its influence was negligible.
Only in retrospect does one observe that it contained a number of important
ideas whose rediscovery by others had a significant impact on the
development of the subject.  The truly monumental work
{\it{Idealtheorie in Ringbereichen}}  belongs to one of the mainstreams
of abstract algebra, commutative ring theory, and may be regarded as
the first paper in this vast subject...''
Though the terminology -  ideal theory, rings, Noetherian rings, the chain
condition, etc. - is unfamiliar to most physicists, one can read Weyl's
lucid account in his memorial address and gain some
understanding of why Jacobson says `` By now her contributions have become
so thoroughly absorbed into our mathematical culture that only rarely are
they specifically attributed to her.''

In 1924 B. L. van der Waerden came to  G\"ottingen having just finished his
university course at  Amsterdam. According to Kimberling, van der Waerden then
mastered her theories, enhanced them with findings of his own, and like no one
else promulgated her ideas.\cite{kimb}  In her obituary,  van der Waerden wrote
that ``her abstract, nonvisual conceptualizations met with little recognition
at first.  This changed as the productivity of her methods was gradually
perceived even by those who did not agree with them. ... Prominent
mathematicians from all over Germany and abroad came to consult with her and
attend her lectures. ... And today, carried by the strength of her thought,
modern algebra appears to be well on its way to victory in every part of the
civilized world'' \cite{van} His book {\it{Moderne Algebra}}, as is credited on
the title page, is based on the  lectures of Emmy Noether and Emil Artin.
According to Garrett Birkhoff, this book precipitated a revolution in the
history of algebra. `` Both the axiomatic approach and much of the content of
`modern' algebra dates back to before 1914.   However, even in 1929 its
concepts and methods were still considered to have marginal interest as
compared with those of analysis... By exhibiting their mathematical and
philosophical unity, and by showing their power as developed by Emmy Noether
and her younger colleagues (most notably E. Artin, R. Brauer and H. Hasse), van
der Waerden made `modern algebra' suddenly seem central in mathematics. It is
not too much to say that the freshness and enthusiasm of his exposition
electrified the mathematical world.''\cite {birk} The first edition of
 {\it{Moderne Algebra}} was published in 1931.   In the 1950's when  I was a
 graduate
student in the University of Chicago, modern algebra certainly appeared central
to us. Though we were graduate students
studying physics,   modern algebra was a subject we all
aspired to learn. I believe it affected profoundly how modern physicists think
and work.

The major papers in the third and final period, 1927-1935,  are
{\it{Hyperkomplexe Gr\"ossen und Darstellungstheorie}} (no.33),  {\it{Beweis
eines Hauptsatzes in der Theorie der Algebren}} (no.38), and  {\it{
Nichtkommutative Algebren}} (no.40). The reader is refered to Jacobson [3] for
a detailed description of their content and significance from a contemporary
point of view.   Weyl says about the work of this period that ``The  theory of
non-commutative algebras and their representations was built up by Emmy Noether
in a new unified, purely conceptual manner by making use of all the results
that had been accumulated by the ingenious labors of decades by Frobenius,
Dickson, Wedderburn and others.''  She found the idea of automorphism useful,
and made major contributions to  cohomology theory. The work of this  period is
of great interest to present-day mathematicians, and  theorists are finding it
of value in their analyses of quantum field theories \cite{tomb} and lattice
 gauge
field theories \cite{dhok}. It is also important, for example, in modern number
theory.  According to Jacobson, ``of equal importance with  [her] specific
achievements were Noether's contributions in unifying the field and providing
the proper framework for future research.''

     According to Weyl of her predecessors in algebra and number theory,
Richard Dedekind was most closely related to her. She edited
with Fricke and Ore the collected mathematical works of Dedekind, and the
commentaries are mostly hers.
She also edited the  correspondence of Georg Cantor and  Richard Dedekind.
 In addition to
doing  mathematics, giving lectures and lecture courses,
supervising doctoral students and writing papers, Emmy Noether
was a voluminous
correspondent, especially with Ernst Fischer, a successor to Gordan
 in Erlangen; and H. Hasse, and was
very active editing for Mathematische  Annalen. \cite{Dick}

The following tribute  to Noether's work was written by A. Einstein.
 ``In the realm of algebra, in which the most gifted mathematicians have been
busy for centuries, she discovered methods which have proved of enormous
importance... Pure mathematics is, in its way, the poetry of logical ideas.
One seeks the most general ideas of operation which will bring together in
simple, logical and unified form the largest possible circle of formal
relationships.  In this effort toward logical beauty spiritual formulas are
discovered necessary for the deeper penetration into the laws of nature.''
\cite{eins}

\section{ BRIEF BIOGRAPHY }

Emmy Noether was born Amalie Emmy Noether in Erlangen, Germany in 1882.
Her father Max was a professor
of mathematics in the university. She was born into a mathematical family.
There were people of known mathematical ability on her grandmother's side,
and her younger brother Fritz became an applied mathematician.
Because both father and
daughter published papers frequently referred to in the mathematical
literature, the work done by Max is sometimes confused with that of his
daughter. Max was a distinguished mathematician best known
for the papers he published  in 1869 and 1872. \cite{noet2} This work
was important for
the development of algebraic geometry; it proved
what the mathematicians
call Noether's fundamental theorem,  or the
residue theorem.  The theorem specifies conditions under which a given
polynomial F(x,y) can be written as a linear combination of two given
polynomials f and g with polynomial coefficients.  Hermann Weyl
says about Max's work
``...Clebsch had introduced Riemann's ideas into the geometric theory of
algebraic curves and Noether became, after Clebsch had passed away young, his
executor in this matter: he succeeded in erecting the whole structure of the
algebraic geometry of curves on the basis of the so-called Noether residual
theorem.'' \cite{weyl}
About the man he said ``... such is the impression I gather from his papers and
even more from the many obituary biographies [ he wrote ] .... a very
intelligent, warm-hearted harmonious man of many-sided interests and sterling
education.''
Max was successor to Felix Klein. Klein made Erlangen famous by
announcing  the Erlangen
Program while he was professor there.  The Erlangen Program
 was to  classify and study geometries according to properties
which remain invariant under appropriate transformation groups. With this
progam ``various geometries previously studied separately were put under one
unifying theory which today still serves as a guiding principle in
geometry.''\cite{kimb} Klein left to join Hilbert in G\"ottingen,
and Max Noether and
 Paul Gordan   were the two Erlangen professors   mainly responsible
for the mathematical atmosphere in which Emmy grew up.
Little has been written so far about Emmy's mother.

During most  of the 19th century women were not allowed in European and North
American universities and laboratories.  The formal education of girls ended at
age fourteen in Germany.  However, as Emmy was growing up change was in the
air. In 1898 the Academic Senate in  the University of Erlangen declared that
the admission of women students would ``overthrow all academic order.'' [2]
Nevertheless in 1900 Emmy got  permission to attend lectures. The university
registry shows then that two of 986 students  attending lectures were female.
However, women were not allowed to matriculate. Emmy attended lectures, and
passed matura examinations at a nearby Gymnasium in 1903.  In the winter she
went to  G\"ottingen and attended lectures given by Schwarzschild, Minkowski,
Klein,  and Hilbert. Of course she was not allowed to enroll. In 1904 it
became possible for females to  enroll in the University of Erlangen and take
examinations with the same rights as male students. She returned and  did a
doctoral thesis under the supervision of her father's friend and colleague Paul
Gordan.  The title  of her thesis {\it{On Complete Systems of Invariants for
Ternary Biquadratic Forms}}. It contains a tabulation of  331 ternary quartic
covariant forms! It was officially registered in 1908.
She was Gordan's only doctoral student. \cite{Dick} She quickly moved on from
 this
calculational phase to  David Hilbert's  more abstract approach  to the theory
of invariants.  In a famous paper of 1888, Hilbert gave a proof by
contradiction of the existence of a finite basis for certain invariants.  It
was the solution to a problem Gordan had worked on for many years and Gordan,
after reading it, exclaimed, ``Das ist nicht Mathematik; das ist Theologie.''
Gordan was an algebraist of the old school.

After obtaining her doctorate, Emmy Noether stayed  in Erlangen in an unpaid
capacity doing her own research, supervising doctoral students  and
occasionally substituting for her father at his lectures  until 1915 when
Hilbert invited her to join his team in G\"ottingen.  This was the most active
and distinguished center of mathematical
research in Europe.  However the mathematics faculty led by Hilbert and
Klein  found it impossible to obtain a university Habilitation for Emmy.
Without  that she could not teach or even give any University lectures. Her
mathematical colleagues all supported her but at that time the Habilitation was
awarded only to male candidates and Hilbert could not get around this.
{}From 1916 to 1919, when finally she was given Habilitation,
she often gave lectures which formally were Hilbert's; the
lectures were advertised as {\it{Mathematisch-Physikalisches Seminar,}}
[ title ], Professor Hilbert with the assistance of Frl. Dr. E. Noether.
Finally awarded Habilitation, she
could  announce her own lectures. She remained, however, in an unpaid position,
and it was not until 1923, when she
was 41, that she was given a university position - but only that of
{\it{nicht-beamteter ausserordentlicher  Professor}}. The  position carried
with
 it
no salary. However, Hilbert was able to arrange for her to have a
{\it{Lehrauftrag}} for algebra which carried a small stipend.

 In 1933 when the Nazi Party came to power,  Jews were forced out of
their academic positions by decree. The Nazis didn't want `Jewish science'
taught in the University. Emmy Noether was a Jewish woman and lost
her position. At that time 3 of the 4 institutes of mathematics
and physics were headed by Jews - Courant, Franck, and Born.  They all
had to leave their teaching positions. Hermann Weyl took over from
Courant for a while thinking he could hold things together, that this was
a transitory bad patch and that reason would prevail.  Before a year was out
he saw otherwise and also left G\"ottingen.  He says of that period:
``A stormy time of struggle like this one we
spent in G\"ottingen in the summer of 1933 draws people closely together; thus
I have a vivid recollection of these months.  Emmy Noether - her courage, her
frankness, her unconcern about her own fate, her conciliatory spirit - was in
the midst of all the hatred and meanness, despair and sorrow surrounding us, a
moral solace.'' Otto Neugebauer's photo of her at the railroad station
leaving G\"ottingen in 1933 is shown here.

There  were only two positions offered Noether in 1933 when she had to flee the
Nazi's. One was a visiting professorship  at Bryn Mawr  supported, in part, by
Rockefeller Foundation funds; and  the other was in Somerville College, Oxford,
where  she was  offered  a stipend of fifty pounds aside from  living
accomodations.  She  went to Bryn Mawr.  While there she was invited to give  a
weekly course of  two hour lectures  at the Institute for Advanced Study in
Princeton.   She traveled there by train each week to do so. Jacobson
attended those lectures in 1935  and recollects that she announced a brief
recess in her course because she had to undergo some surgery.  Apparently the
operation was followed by a virulent infection and she died quite unexpectedly.
According to Weyl,  ``She was at the summit of her mathematical creative
power'' when she died.

Many people have written about how
helpful and influential she was in the work of others. She
not infrequently tended not to have her name included as author
on papers to which she had contributed
in order to promote the careers of
younger people.  She apparently was quite content with this and
didn't feel a necessity to promote her own fame.  She  lived a very simple
life and is reported to have been quite a happy person though she existed
on meager funds. Einstein wrote this tribute to her in his
Letter to the Editor of the New York Times .\cite{eins}
`` The efforts of most human beings are consumed in the struggle for
their daily bread, but most of those who are, either through fortune
or some special gift, relieved of this struggle are largely absorbed in
further improving their worldly lot.  Beneath the effort directed
toward the accumulation of worldly goods lies all too frequently the
illusion that this is the most substantial and desirable end to be
achieved; but there is, fortunately, a minority composed of those
who recognize early in their lives that the most beautiful and satisfying
experiences open to humankind are not derived from the outside, but are
bound up with the development of the individual's own feeling,
thinking and acting.  The genuine artists, investigators and thinkers
have always been persons of this kind.  However inconspicuously the life of
these individuals runs its course, none the less the fruits of their
endeavors are the most valuable contributions which one generation can
make to its successors.''

\acknowledgements

It is a pleasure to express my gratitude to
the organizers of the Conference for providing the occasion to
write this paper, to Basil Gordon for illuminating mathematical discussions and
close editing of the contents particularly
as they deal with mathematics, and to Terry Tomboulis  for helpful
conversations.

\tighten
\twocolumn

\appendix
\section{PUBLICATION LIST}
This list does not contain edited, and annotated, books and papers.
\begin{enumerate}
\scriptsize
\item{} \"Uber die Bildung des Formensystems der tern\"aren
biquadratischen Form.
  Sitz. Ber. d. Physikal.-mediz. Soziet\"at in Erlangen 39
 (1907), pp. 176-179.
\item{}  \"Uber die Bildung des Formensystems der tern\"aren
biquadratischen Form.
   Journal f. d. reine u. angew. Math. 134 (1908),
 pp. 23-90.
\item{} Zur Invariantentheorie der Formen von n Variabeln.
  J. Ber. d. DMV 19 (1910), pp. 101-104.
\item{4.} Zur Invariantentheorie der Formen von n Variabeln.
  Journal f. d. reine u. Angew. Math. 139 (1911), pp. 118-154.
\item{} Rationale Funktionenk\"orper.
J. Ber. d. DMV 22 (1913). pp. 316-319.
\item{} K\"orper und Systeme rationaler Funktionen.
  Math. Ann. 76 (1915), pp. 161-196.
\item{} Der Endlichkeitssatz der Invarianten endlicher Gruppen.
  Math. Ann. 77 (1916), pp. 89-92.
\item{} \"Uber ganze rationale Darstellung der Invarianten
 eines Systems von beliebig vielen Grundformen.
  Math. Ann. 77 (1916), pp. 93-102. (cf. No. 16).
 \item{} Die allgemeinsten Bereiche aus ganzen
 transzendenten Zahlen.
  Math. Ann. 77 (1916), pp. 103-128. (cf. No. 16)
 \item{} Die Funktionalgleichungen der isomorphen Abbildung.
  Math. Ann. 77 (1916), pp. 536-545.
 \item{} Gleichungen mit vorgeschriebener Gruppe.
  Math. Ann 78 (1918), pp. 221-229. (cf. No. 16).
 \item{} Invarianten beliebiger Differentialausdr\"ucke.
  Nachr. v. d. ges. d. Wiss. zu G\"ottingen 1918.
   pp. 37-44.
 \item{} Invariante Variationsprobleme.
  Nachr. v. d. Ges. d. Wiss. zu G\"ottingen 1918, pp. 235-257.
 \item{} Die arithmetische Theorie der algebraischen Funktionen
 einer Ver\"anderlichen
  in ihrer Beziehung zu den \"ubrigen Theorien und zu der
 Zahlk\"orpertheorie.
  J. Ber. d. DMV 28 (1919), pp. 182-203.
 \item{} Die Endlichkeit des Systems der ganzzahligen Invarianten
 bin\"arer Formen.
  Nachr. v. d. Ges. d. Wiss. zu G\"ottingen 1919, pp. 138-156.
 \item{} Zur Reihenentwicklung in der Formentheorie.
  Math. Ann. 81 (1920), pp. 25-30.
\item{}Moduln in nichtkommutativen Bereichen,
 insbesondere aus Differential-
  und Differen-zenaus-dr\"ucken.
 Co-authored by W. Schmeidler. Math.
 Zs. 8 (1920), pp. 1-35.
 \item{} \"Uber eine Arbeit des im Kriege gefallenen K. Hentzelt zur
 Eliminationstheorie.
  J. Ber. d. DMV 30 (1921), p. 101
 \item{} Idealtheorie in Ringbereichen
  Math. Ann. 83 (1921), pp. 24-66.
 \item{} Ein algebraisches Kriterium f\"ur absolute Irreduzibilit\"at.
  Math Ann. 85 (1922), pp. 26-33.
 \item{} Formale Variationsrechnung und
  Differentialinvarianten.
  Encyklop\"adie d. math. Wiss. III, 3 (1922), pp. 68-71
(in: R. Weitzenb\"ock, Differentialinvarianten).
 \item{} Bearbeitung von K. Hentzelt: Zur Theorie der
 Polynomideale und Resultanten.
  Math. Ann. 88 (1923), pp. 53-79.
 \item{} Algebraische und Differentialvarianten.
  J. Ber. d. DMV 32 (1923), pp. 177-184.
 \item{} Eliminationstheorie und allgemeine Idealtheorie.
  Math. Ann. 90 (1923), pp. 229-261.
 \item{} Eliminationstheorie und Idealtheorie.
  J. Ber. d. DMV 33 (1924), pp. 116-120.
 \item{} Abstrakter Aufbau der Idealtheorie im algebraischen Zahlk\"orper.
  J. Ber. d. DMV 33 (1924), p. 102.
 \item{} Hilbertsche Anzahlen in der Idealtheorie.
  J. Ber. d. DMV 34 (1925), p. 101
 \item{} Gruppencharaktere und Idealtheorie.
  J. Ber. d. DMV 34 (1925), P. 144.
 \item{} Der Endlichkeitssatz der Invarianten endlicher
 linearer Gruppen der Charakteristik p.
  Nachr. v. d. Ges. d. Wiss. zu  G\"ottingen 1926,
 pp. 28-35.
 \item{}Abstrakter Aufbau der Idealtheorie in algebraischen
  Zahl-und Funktionenk\"orpern.
  Math. Ann. 96 (1927), pp. 26-61.
 \item{} Der Diskriminantensatz f\"ur die Ordnungen eines
 algebraischen Zahl - oder Funktionenk\"orpers.
  Journal f. d. reine u. angew. Math. 157 (1927),
 pp. 82-104.
 \item{} \"Uber minimale Zerf\"allungsk\"orper irreduzibler Darstellungen.
 Co-authored by
  R. Brauer. Sitz. Ber. d.Preuss. Akad. d. Wiss. 1927, pp. 221-228
 \item{} Hyperkomplexe Gr\"oss en und Darstellungstheorie in
 arithmetischer Auffassung.
  Atti Congresso Bologna 2 (1928), pp. 71-73.
 \item{} Hyperkomplexe Gr\"ossen und Darstellungstheorie.
  Math. Zs. 30 (1929), pp. 641-692.
 \item{} \"Uber Maximalbereiche aus ganzzahligen Funktionen.
  Rec. Soc. Math. Moscou 36 (1929), pp. 65-72.
 \item{} Idealdifferentiation und Differente.
  J. Ber. d. DMV 39 (1930), p. 17.
 \item{} Normalbasis bei K\"orpern ohne h\"ohere Verzweigung.
  Journal f. d. reine u. angew. Math. 167 (1932),
 pp. 147-152.
 \item{} Beweis eines Hauptsatzes in der Theorie der Algebren.
 Co-authored by R. Brauer and
  H. Hasse. Journal f. d.  reine u. angew. Math. 167 (1932), pp. 399-404.
 \item{} Hyperkomplexe Systeme in ihren Beziehungen zur
 kommutativen Algebra und zur Zahlentheorie.
  Verhandl. Intern. Math.-Kongre\ss Z\"urich 1 (1932),
 pp. 189-194.
 \item{} Nichtkommutative Algebren.
  Math. Zs. 37 (1933), pp. 514-541.
 \item{} Der Hauptgeschlechtssatz f\"ur relativ-galoissche
Zahlk\"orper.
  Math. Ann. 108 (1933), pp. 411-419.
 \item{} Zerfallende verschr\"ankte Produkte und ihre
 Maximalordnungen.
  Actualit$\acute{e}$s scientifiques et industrielles 148 (1934)
 (15 pages).
 \item{} Idealdifferentiation und Differente.
  Journal f. d. reine u. angew. Math. 188 (1950), pp. 1-21.

\end{enumerate}

\onecolumn

\section{Excerpt from habilitation submission.}

\baselineskip=20pt

Noether characterized  her
published papers from the period 1907 to 1918 in her submission for
 Habilitation. The submission   reads
in part (number insertions refer to the list of publications in Appendix A):

 ``My dissertation and a later paper ... belong to the theory of formal
invariants, as was natural for me as a student of Gordan. The lengthiest paper,
`Fields and Systems  of Rational Functions' (6) concerns questions about
general bases; it completely solves the problem of rational representation and
contributes to the solution of other finiteness problems.  An application of
these results is contained in `The Finiteness Theorem for Invariants of Finite
Groups' (7) which offers an absolutely elementary proof by actually  finding a
basis. To this line of investigation also belongs the paper `Algebraic
Equations with Prescribed Group' (11) which is a contribution to the
construction of such equations for any field range....  The paper `Integral
Rational Representation of Invariants' (8) proves valid a conjecture of D.
Hilbert ...  With these wholly algebraic works belong two additional works ....
'A Proof of finiteness for Integral Binary Invariants' (15) ... and an
investigation with W. Schmeidler of noncommutative one-sided modules...
`Alternatives with Nonlinear Systems of Equations'...  The longer work `The
Most General Ranges of Completely Transcendental Numbers' (9) uses along with
algebraic and number-theoretic techniques some abstract set theory ...In this
same direction is the paper `Functional Equations and Isomorphic Mapping'  (10)
which yields the most  general isomorphic mapping of an arbitrarily abstractly
defined field.  Finally, there are two works on differential invariants and
variation problems (12,13)...'' ~\cite{Dick}

\tighten
\twocolumn

\section{Titles from a recent issue of Current Contents}

 \begin{enumerate}
\scriptsize
\item{} V. Iyer and R.M. Wald,
    {\it Some Properties of the Noether Charge and a Proposal for
Dynamical Black Hole Entropy}, Phys. Rev. D {\bf 50}, 846, {1994}.

\item{}
   O. Castanos, R. Lopezpena  and V. I. Manko, {\it
      Noether Theorem and Time-dependent Quantum Invariants}, J. Phys.
A: Mathematical and General, {\bf 27}, 1751 (1994).

\item{}
    M. Forger, J. Laartz and U. Schaper, {\it The Algebra of the
Energy-momentum Tensor and the Noether Currents in Classical Non-linear
Sigma Models}, Commun. Math. Phy. {\bf 159}, 319
(1994).

\item{}
   P. G. Henriques,
     {\it The Noether Theorem and the Reduction Procedure for the
Variational Calculus in the Context of Differential Systems}, Comptes
Rendus De L Academie Des Sciences Serie I-Mathematique {\bf 317}, 987
(1993).

\item{}
    R. M. Wald, {\it Black Hole Entropy is the Noether Charge}, Phys.
Rev. D {\bf 48} N8:R3427 (1993).

\item{}
    Z. P. Li, {\it
      Generalized Noether Theorem and Poincare-Cartan Integral Invariant
for Singular High-order Langrangian in Fields Theories}, Science in
China Series-A-Mathematics Physics Astronomy Technological Sciences {\bf
36}, 1212 (1993).

\item{}
    D. E. Neuenschwander and S. R. Starkey, {\it Adiabatic Invariance
Derived from the Rund-Trautman Identity and Noether Theorem}, Am.
J.  Phys. {\bf 61}, 1008 (1993).

\item{}
    L. Lusanna, {\it
      The Shanmugadhasan Canonical Transformation, Function Groups and
the Extended 2nd Noether Theorem}, International Journal of Modern
Physics A, {\bf 8}, 4193 (1993).

\item{}
   A. Weldo, {\it Hard Thermal Loops and their Noether Currents}, Can.
J. Phys. {\bf 71}, N5-6:300 (1993).

\item{}
    A. Milinski, {\it Skolem-Noether Theorems and Coalgebra Actions},
Communications in Algebra {\bf 21}, 3719 (1993).

\item{}
    Z. P. Li, {\it Noether Theorem and its Inverse Theorem in Canonical
Formalisam for Nonholonomic Nonconservative Singular System}, Chinese
Science Bulletin {\bf 38}, 1143 (1993).

\item{} F. M. Mahomed, A. H. Kara and P. G. L. Leach, {\it
Lie and Noether Counting Theorems for One-Dimensional Systems}, Journal
of Mathematical Analysis and Applications  {\bf 178} 116 (1993).
    N1:116-129.

\item{} G. Reinish and J. C. Fernandez, {\it Noether Theorem and
the Mechanics of Nonlinear Solitary Waves}, Phys. Rev. B: Condensed
Matter {\bf 48} 853 (1993).

 \item{} A. I. Tuzik,  {\it On the Noether Conditions of One Dual Discreet
Equation of Convolution Type with almost Stabilizing Multipliers},
Dokl. Akad. Nauk Belarusi, {\bf 37} 118 (1993).

\item{} V. A. Dorodnitsyn {\it The Finite-Difference Analogy of
Noether Theorem}, Dokl. Akad. Nauk, {\bf 328} 678 (1993).

\item{} M. Forger and J. Laartz {\it The Algebra of the
Energy-Momentum Tensor and the Noether Currents in Off-critical WZNW
Models}, Modern Physics Letters A., {\bf 8} 803 (1993).

\item{}
     X. C. GAO, J. B. Xu, T. Z. Qian and J. Gao {\it
      Quantum Basic Invariants and Classical Noethern Theorem},
      Physica Scripta, {\bf 47} 488 (1993).

\item{}
    N. A. Lemos {\it Symmetries, Noether Theorem and Inequivalent
Lagrangians Applied to Nonconservative Systems}, Revista Mexicana De
Fisica, {\bf 39} 304 (1993).

\item{}
    J. L. Colliotthelene {\it
      The Noether-Lefschetz Theorem and Sums of 4 Squares in the
Rational Function Field R(X,Y)}, Compositio Mathematica, {\bf 86} 235
(1993).

\item{}
  S. Caenepeel {\it Computing The Brauer-Long Group of a Hopf Algebra.
2.  The Skolem-Noether Theory}, Journal of Pure and Applied Algebra,
{\bf 84} 107 (1993).

  \item{}
 Z. P. Li {\it Generalized Noether Theorems in Canonical Formalism
for Field Theories and their applications}, Int. J. Theor. Physi.
{\bf 32} 201 (1993).

\item{}
O. Castanos and R. Lopezpena {\it Noether Theorem and Accidental
Degeneracy}, J. Physi A: Mathematical and General {\bf 25} 6685 (1992).

\item{} X. Gracia and J. M. Pons {\it A Hamiltonian Approach to Lagrangian
Noether Transformations}, J. Physi. A:  Mathematical and General {\bf 25}
6357 (1992).

\item{} K. Kurano {\it Noether Normalizations for Local Rings of Algebraic
Varieties}, Proceedings of American Mathematical Society {\bf 116} 905
(1992).

\item{}  L. B. Szabados {\it On Canonical Pseudotensors, Sparling form and
Noether Currents},  Classical and Quantum Gravity {\bf 9} 2521 (1992).

\item{} E. Gonzalezacosta and M. G. Coronagalindo {\it Noether Theorem and the
Invariants for Dissipative and Driven Dissipative Like Systems}, Revista
Mexicana De Fisica {\bf 38} 511 (1992).
\item{} S. L. Guo {\it Noether (Artin) Classical Quotient Rings of Duo Rings
and Projectivity and Injectivity of Simple Modules over Duo Rings},
Chinese Science Bulletin {\bf 37} 877 (1992).

\item{} X. C. Gao , J. B. Xu and J. Gao {\it A New Approach to the Study of
Noether Invariants and Symmetry Algebras}, Prog. Theor. Phys. {\bf 87}
861 (1992).

\item{} Z. P. Li {\it Noether Theorems in Canonical Formalism and their
Applications}, Chinese Science Bulletin {\bf 37} 258 (1992).

\item{} J. Krause {\it Some Remarks on the Generalized Noether Theory of
Point Summetry Transformations of the Lagrangian}, J. Phys. A:
Mathematical and General {\bf 25} 991 (1992).

\item{} M. Kobayashi {\it On Noether Inequality for Threefolds}, Journal of
the Mathematical Society of Japan {\bf 44} 145 (1992).

\item{}  S. K. Luo {\it Generalized Noether Theorem for Variable Mass
Higher-order Nonholonomic Mechanical Systems in Noninertial Reference
Frames}, Chinese Science Bulletin {\bf 36} 1930 (1991).

\item{}  G. Dzhangibekov {\it On the Conditions of Noether Property and Index
of some 2-Dimensional Singular Integral Operators}, Dokl. Akad. Nauk
SSSR {\bf 319} 811 (1991).

\item{}  R. D. Sorkin {\it The Gravitational Electromagnetic Noether Operator
and the 2nd-Order Energy Flux}, Proc. R. Soc. London Ser. A. {\bf 435}
635 (1991).

\item{} C. Cilibeto and A. F. Lopez {\it On the Existence of Components of
the Noether-Lefschetz Locus with Given Codimension}, Manuscripta
Mathematica {\bf 73} 341 (1991).

\item{} J. F. Carinena. Fernandeznunez and E. Martinez {\it A Geometric
Approach to Noether 2nd Theorem in Time-Dependentr Lagrangian
Mechanics}, Letters in Mathematical Physics {\bf 23} 51 (1991).

\item{}  V. G. Kravchenko and G. S. Litvinchuk {\it Noether Theory of
Unbounded Singular Integral Operators with a Carleman Shift},
Mathematical Notes {\bf 49} 45 (1991).

\item{}  M. Koppinen {\it A Skolem-Noether Theorem for Coalgebra
Measurings}, Archiv Der Mathematik {\bf 57} 34 (1991).

\item{} I. C. Moreira and M. A. Almeida {\it Noether Symmetries and the
Swinging Atwood Machine}, Journal De Physique II. {\bf 1} 711 (1991)

\item{}  B. Y. Shteinberg, {\it The Noether Property and the Index of
Multidimensional Convolutions with Coefficients of Fast Oscillating
Type}, Siberian Mathematical Journal {\bf 31} 682 (1990).

\item{}  L. Lusanna {\it The 2nd Noether Theorem as the Basis of the Theory
of Singular Lagrangians and Hamiltonians Constraints}, Rivista Del Nuovo
Cimento {\bf 14} 1 (1990).

\item{}  S. O. Kim {\it Noether-Lefschetz Locus for Surfaces}, Transactions
of the American Mathematical Society {\bf 324} 369 (1991).

\item{} H. A. Alkuwari and M. O. Taha {\it Noether Theorem and Local Gauge
Invariance}, Am. J. Phys. {\bf 59} 363 (1991).

\item{}  F. A. Lunev {\it Analog of Noether Theorem for Non-Noether and
Nonlocal Symmetries}, Theoretical and Mathematical Physics {\bf 84} 816
(1990).

\item{} Z. P. Li and X. Li {\it Generalized Noether Theorems and
Applications}, Int. J. Theor. Phys. {\bf 30} 225 (1991).

\item{} H. Aratyn E. Nissimov and S. Pacheva {\it Infinite-Dimensional
Noether Symmetry Groups and Quantum Effective Actions from Geometry},
Phys. Lett B.  {\bf 255} 359 (1991).

\item{} J. Stevens {\it A Generalisation of Noether Formula for the Number of
Virtual Double Points to Space Curve Singularities}, Archiv der
Mathematik {\bf 56} 96 (1991).

\item{}  A. F. Lopez {\it Noether-Lefschetz Theory and the Picard Group of
Projective Surfaces},  Memoirs of the American Mathematica Society, {\bf
89} 438 (1991).

\item{}  O. Castanos, A. Frank and R. Lopezpena {\it Noether Theorem and
Dynamical Groups in Quantum Mechanics}, Journal Phy. A: Mathematical
and General {\bf 23} 5141 (1990).

\item{}  G. Profilo and G. Soliani {\it Noether Invariants and Complete
Lie-Point Symmetries for Equations of the Hill Type}, Prog. of Theor.
Phys. {\bf 84} 974 (1990).

\item{} J. D. McCrea, F. W. Hehl and E. W. Mielke {\it Mapping Noether
Identities into Bianchi Identities in General Relativistic Theories of
Gravity and in the Field Theory of Static Lattice Defects},
Int. J.  Theor. Physi. {\bf 29} 1185 (1990).

\item{} J. F. Zhang {\it Generalized Noether Theorem of High-Order
Nonholonomic Nonconservative Dynamical Systems},  Chinese Science Buletin
{\bf 35} 1758 (1990).

\item{}  M. Beattie and K. H. Ulbrich {\it A Skolem-Noether Theorem for Hopf
Algebra Actions}, Communications in Algebra {\bf 18} 3713 (1990).

\item{}  C. Ferrario and A. Passerini {\it Symmetries and Constants of
Motion for Constrained Lagrangian Systems - A Presymplectic Version of
the Noether Theorem}, J. Physi. A:  Mathematical and General {\bf 23}
5061 (1990).

\item{} O. Debarre and Y. Laszlo {\it Noether-Lefschetz Locus for Abelian
Varieties}, Comptes Rendus De L. Academie Des Sciences Series
I-Mathematique {\bf 311} 337 (1990).

\item{} H. Zainuddin {\it Noether Theorem in Nonlinear Sigma-Models with a
Wess-Zumino Term}, J. Math. Phy. {\bf 31} 2225 (1990).

\item{}  Z. P. Li and X. Li {\it Generalized Noether Theorem and Poincare
Invariant for Nonconservative Nonholonomic Systems}, Int. J. Theor. Phy.
{\bf 29} 765 (1990).

\item{}  C. Voisin {\it Noether-Lefschetz Locus in 6th and 7th Degrees},
Compositio Mathematica {\bf 75} 47 (1990).

\item{}  I. C. Moreira {\it On Noether Theorem for  Time-Dependent
Oscillators - Comment}, Europhysics Letters {\bf 12} 391 (1990).

\item{}  H. Aratyn E. Nissimov, S. Pacheva and A. H. Zimerman {\it The
Noether Theorem for Geometric Actions and Area Preserving
Diffeomorphisms on the Torus}, Phys. Lett. B {\bf 242} 377 (1990).

\item{} J. Castellanos {\it Hamburger-Noether Matrices over Rings}, Journal
of Pure and Applied Algebra {\bf 64} 7 (1990).

\item{}  Y. I. Karlovich and I. M. Spitkovskii {\it Factorization of Almost
Periodic Matrix-Valued Functions and the Noether Theory for Certain
Classes of Equations of Convolution Type}, Mathematics of the
USSR-Izvestiya {\bf 53} 281 (1989).

\item{}  X. Wu {\it On a Conjecture of Griffiths-Harris Generalizing the
Noether-Lefschetz Theorem}, Duke Mathematical Journal {\bf 60} 465
(1990).

\item{}  I. Y. Krivskii and V. M. Simulik {\it Noether Analysis of Zilch
Conservation Laws and Their Generalization for the Electromagnetic
Field 2; Use of Poincare-Invariant Formulation of the Principle of Least
Action}.  Theoretical and Mathematical Physics {\bf 80} 912 (1989).

\item{}  I. Y. Krivskii and V. M. Simulik {\it Noether Analysis of Zilch,
Conservation Laws and their Generalization for the Electromagnetic Field
1; Use of Different Formulations of the Principle of Least Action},
Theoretical and Mathematical Physics {\bf 80} 864 (1989).

\item{} D. A. Cox {\it The Noether-Lefschetz Locus of Regular Elliptic
Surfaces with Section and PG Greater-Than-or-Equal-to 2}, American
Journal of Mathematics {\bf 112} 289 (1990).

\item{} D. L. Karatas and K. L. Kowalski {\it Noether Theorem for Local Gauge
Transformations}, Am. J. Physics {\bf 58} 123 (1990).

\item{}  Z. Y. Gu and S. W. Qian {\it Noether Theorem Invariants for a
Time-Dependent Damped Harmonic Oscillator with a Force Quadratic in
Velocity}, Europhysics Letters, {\bf 10} 615 (1989).

\item{}  C. Voisin {\it Small Dimension Components of the Noether-Lefschetz
Locus}, Commentarii Mathematici Helvetici {\bf 64} 515 (1989).

\item{} L. Moretbailly {\it The Noether-Formula for Arithmetic Surfaces},
Inventiones Mathematicae {\bf 98} 491 (1989).

\item{} T. Maeda {\it Noether Problem for A5}, Journal of Algebra {\bf 125}
418 (1989)

\item{} E. W. Mielke, F. W. Hehl and J. D. McCrea {\it Belinfante Type
Invariance of the Noether Identities in a Riemannian and a Weitzenbock
Spacetime}, Phy. Lett. A {\bf 140} 368 (1989)
\end{enumerate}

\onecolumn

\end{document}